\documentclass[reprint,amsmath,amssymb,aps,noeprint]{revtex4-2}
\usepackage{graphicx}
\usepackage{dcolumn}
\usepackage{bm}
\usepackage{hyperref}
\usepackage{braket}
\usepackage{mathtools}
\usepackage[usenames,dvipsnames]{xcolor}

\hypersetup{
    colorlinks,
    citecolor=blue,
    linkcolor=blue,
    urlcolor=blue
}

\usepackage{soul}
\usepackage[whole]{bxcjkjatype}

\begin{document}

\preprint{APS/123-QED}

\title{Uncertainty relation for indirect measurement}

\author{Ryo Mihashi}
\email{mihashi@biom.t.u-tokyo.ac.jp}
\affiliation{Department of Information and Communication Engineering, Graduate School of Information Science and Technology, The University of Tokyo, Tokyo 113-8656, Japan}

\author{Yoshihiko Hasegawa}
\email{hasegawa@biom.t.u-tokyo.ac.jp}
\affiliation{Department of Information and Communication Engineering, Graduate School of Information Science and Technology, The University of Tokyo, Tokyo 113-8656, Japan}

\date{\today}
\begin{abstract}
    Indirect measurement can be used to read out the outcome of a quantum system without resorting to a straightforward approach, and it is the foundation of the measurement uncertainty relations that explain the incompatibility of conjugate observables. While measurement uncertainty has been established for the accuracy of quantum measurement, thermodynamic uncertainty is being researched as a new perspective on the accuracy of fluctuating observables that trades off a physical cost in the classical or quantum regime. Following these ideas, we derive a new measurement uncertainty with respect to indirect measurement in the light of quantum thermodynamics. The obtained result shows a reciprocal relation between a survival activity, which is a frenetic contribution to the thermodynamic system, and noise due to the measurement approach. Moreover, this uncertainty can also affect the disturbance of an incompatible counterpart through the universal noise-disturbance relation.
\end{abstract}

\maketitle

\section{\label{sec:1}Introduction}

Indirect measurement is adopted as a beneficial method when we cannot directly access an observable in quantum systems of interest. Theoretically, the method is constructed with an ancillary system, which is often called a probe, and a suitable observable. Since measurement of a quantum system inevitably demolishes its state as a quantum nature, we can design alternative measurement steps such that only the attached probe system will be destroyed, while keeping the state of the principal system intact \cite{Grangier1998}.

Indirect measurement has also provided a way to tackle the incompatibility of conjugate observables. We cannot simultaneously perform a projective measurement of incompatible observables defined in the same quantum system without producing an error. In principle, accurately measuring one observable disturbs the current quantum state, and therefore its conjugate is unavoidably influenced by the post-measurement state, which may lead to unfavorable results. Instead, we can choose a joint measurement, another framework that allows us to simultaneously measure such conjugate observables with an ancillary system and several commuting observables of the entire system, in a similar way to the indirect method. For example, joint measurement has been employed in quantum optics when we want to know the quadrature amplitudes of squeezed coherent states of a photon \cite{Weedbrook2012}, especially in heterodyne detection \cite{Yamamoto1986, Leonhardt1993}.

Joint measurement not only gives us another choice but also quantifies the non-commutativity of the observables in the principal system. Other than the Heisenberg uncertainty principle deduced by the postulates of quantum mechanics, the measurement uncertainty shows a complementary relation with regard to the measurement of incompatible observables thanks to this setting, and much research has been dedicated to this problem \cite{Arthurs1965, Arthurs1988, Ozawa2003, Watanabe2011, Fujikawa2012, Branciard2013, Lu2014}. 

Representatively, for arbitrary observables $A$ and $B$, Ozawa's relation provides a universal limitation on quantum measurements:
\begin{equation}
  \Delta N_A \Delta N_B + \Delta A \Delta N_B + \Delta N_A \Delta B \geq \dfrac{1}{2} | \langle [A, B]\rangle | ,
  \label{eq:Ozawa}
\end{equation}
where $\Delta (\cdots)$ and $\langle \cdots \rangle$ denote the standard deviation and the expectation value, respectively, of the observable with respect to a quantum state, and the operators $N_A$ and $N_B$ are noise operators that are defined by the subtraction of the approximate observables from the original ones to quantify their deviations [cf. Eq.~\eqref{eq:N_op_def}]. Besides Eq.~\eqref{eq:Ozawa}, the Arthur-Kelly-Goodman relation \cite{Arthurs1965, Arthurs1988} shows an intuitive bound in some limited situations \cite{Ozawa2004}, the tighter relation has been proven \cite{Branciard2013}, and improvement for a relation independent from the probe state has been proposed \cite{Lu2014}.

These relations have always been connected to the disturbance of an observable, namely the back-action of the measurement of non-commuting counterparts, and it is known as the noise-disturbance relation \cite{Ozawa2003,Ozawa2004, Hall2004, Ozawa2004a}.
The measurement uncertainty relations like Eq.~\eqref{eq:Ozawa} quantify the degree of difference induced by the joint measurement of conjugate variables, and importantly, imply how precisely the current measurement can be performed. Hence, the indirect measurement establishes a central clue to explaining these properties.

In addition to the uncertainty relation for joint measurements, a relation between the precision of quantum measurements and thermodynamic costs was recently unveiled, which is referred to as the thermodynamic uncertainty relation (TUR). TURs reflect the extent of accuracy of the measurement of the probabilistic observable $A$, which are usually defined as 
\begin{equation}
  \dfrac{\Delta A^2}{\langle A\rangle^2} \geq \mathcal{P}_{\mathrm{min}}\notag.
  \label{eq:TUR_def}
\end{equation}
Here $\mathcal{P}_\mathrm{min}$ is a lower bound comprising thermodynamic costs, which gets smaller for higher thermodynamic cost. 
Therefore, Eq.~\eqref{eq:TUR_def} shows that less fluctuation can be achieved at the expense of higher thermodynamic costs. 
Typically, for steady-state classical Markov processes, $\mathcal{P}_\mathrm{min}$ is given by $2 / \Sigma$,
where $\Sigma$ is entropy production. 
In other words, the signal-to-noise ratio of $A$ is bounded when the system of interest experiences an irreversible process. TURs have been proven and extended to several classes in classical and quantum systems \cite{Barato2015, Gingrich2016, Hasegawa2018, Dechant2018b, DiTerlizzi2019, Guarnieri2019, VanVu2020, Hasegawa2020QTUR, Hasegawa2021, Horowitz2020}. TURs show their significance especially in quantum regimes. It has been pointed out that TURs in quantum regimes can violate classical counterparts because the quantum coherence of the state \cite{Kalaee2021, Menczel2021} is able to make a contribution as a resource of accuracy.

In this paper, based on Ozawa's configuration, we show a new measurement uncertainty relation for indirect measurement from the viewpoint of thermodynamics. Our result is derived from an information theoretical approach and bestows the thermodynamic viewpoint on the field of measurement uncertainty. The obtained relation shows a trade-off between a survival activity, which is a thermodynamic quantity, and the additional error generated from the indirect approach by bounding them with a coefficient of variation (CV) for the premeasurement state.

\section{\label{sec:2}Indirect Measurement}

In this section, we review the indirect measurement approach. This approach not only provides us with a method to approximately or precisely measure an observable defined in the main system, but also helps to evaluate the error occurring through the method. 
Let S and P be the principal system and the probe system, respectively. The system P is regarded as the apparatus that is exploited to perform the indirect measurement of the system S's observable $A$, which we intend to measure. Let $\rho_{\mathrm{S,P}}$ be the initial quantum state of the composite system $\mathrm{S}+\mathrm{P}$. 
Here, we introduce an approximate observable $\mathcal{A}$, which is defined for the total system $\mathrm{S}+\mathrm{P}$. We suppose that $\mathcal{A}$ has the spectral decomposition
\begin{equation}
  \mathcal{A} = \sum_k r_k P_k,
\end{equation}
where $r_k$ is the $k$th eigenvalue and $P_k$ is the projection operator of $\mathrm{S}+\mathrm{P}$ corresponding to $r_k$. For simultaneous measurements, we select two approximate observables such that they share the common projection operators $\{ P_k \}$. As explained later, these operators are used to indirectly retrieve the genuine observable $A$ \cite{Ozawa2001}.

Practically, approximate observables can be prepared from the interaction between S and P. That is, we measure the composite system after an interaction with a unitary operator $U$ induced by an interaction Hamiltonian. Let us consider the formulation from the Heisenberg picture. To accomplish this measurement, we can employ the meter observable $M$ of $\mathrm{S}+\mathrm{P}$ and the correlation between the quantum states of S and P resulting from this interaction \cite{Ozawa2001, Ozawa2003}. 
The approximate observable $\mathcal{A}$ can be expressed by
\[
  \mathcal{A} := U^{\dagger}MU.
\]
To use this elaborate technique to obtain knowledge about the actual observable $A$, we assume the following unbiasedness condition:
\begin{equation}
    \langle U^{\dagger}MU\rangle = \langle A\rangle,
    \label{eq:agreement_of_mean}
\end{equation}
where $\langle \cdots \rangle$ denotes the ensemble average for the relevant quantum states, and here we average $U^{\dagger}MU$ and $A$ over the initial state on both sides of Eq.~\eqref{eq:unbiased}. The left-hand side of Eq.~\eqref{eq:unbiased} represents the expected value of $M$ after the interaction and the right-hand side does the same but for the initial state of S. Because this agreement of the two mean values should hold on an arbitrary initial state $\rho_\mathrm{S}$, it follows that
\begin{equation}
    A = \mathrm{tr}_\mathrm{P}[U^{\dagger}MU (I_\mathrm{S} \otimes \rho_\mathrm{P})]
    \label{eq:unbiased}
\end{equation}
with the identity $I_\mathrm{S}$. Hereafter, a subscript of $I$ represents the system on which the identity acts. This requirement guarantees that the mean value of $A$, which is a direct measurement of S, should agree with that of $M$. Note that this measurement procedure is not uniquely determined as long as the condition Eq.~\eqref{eq:unbiased} holds.

We can evaluate this measurement procedure by introducing a noise operator \cite{Ozawa2003,Ozawa2004,Ozawa2004a}, defined by the difference between the observable of interest and its meter observable in a Heisenberg representation:
\begin{equation}
    N := U^{\dagger}MU - A\otimes I_{\mathrm{P}}.
    \label{eq:N_op_def}
\end{equation}
The noise operator is accessible and verifiable by experiments \cite{Erhart2012, Rozema2012, Kaneda2014}. When the unbiasedness condition  Eq.~\eqref{eq:unbiased} is satisfied, the mean value of the noise operator approaches zero as $\langle N\rangle=\mathrm{tr}[\rho_\mathrm{S,P}N]=0$. This means that the indirect procedure can recover the actual value $\langle A\rangle$.

We evaluate the difference between the actual measurement and this scheme by the variances of each observable. The unbiasedness of the meter observable Eq.~\eqref{eq:unbiased} immediately leads to
\begin{equation}
  \Delta M^2 = \Delta A^2 + \Delta N^2.
  \label{eq:vardecomp}
\end{equation}
Thus, the variance of the indirect measurement is decomposed into two quantities, the variance of the noise operator $N$ and the actual operator $A$.

As stated in Sec.~\ref{sec:1}, the indirect measurement approach can be exploited to simultaneously measure the incompatible observables $A$ and $B$ satisfying $[A ,B] \neq 0$.  If we want to execute a simultaneous measurement, we introduce commutable meter observables corresponding to the actual observables $A$ and $B$. This helps us to introduce the well-known measurement error uncertainty relations.

\section{\label{sec:3}Quantum Thermodynamic Uncertainty Relation}
TURs describe the relation between a thermodynamic cost and precision \cite{Barato2015, Gingrich2016, Hasegawa2018, Dechant2018b, DiTerlizzi2019, Guarnieri2019, VanVu2020, Hasegawa2020QTUR, Hasegawa2021, Horowitz2020} and mainly means that higher precision requires a larger cost. We incorporated this concept of the trade-off relation into the measurement framework in which the conventional measurement uncertainty has been developed. Here, we briefly visit the thermodynamic uncertainty relation for open quantum systems suitable for unifying these ideas.

Consider an open quantum system composed of a principal system S and an environmental system E. We focus on the fluctuation of the observable $\mathcal{G}$ having the spectral decomposition $\mathcal{G} = \sum_k g_k \ket{g_k}\bra{g_k}$ on E, where $g_k$ is the $k$th eigenvalue of $\mathcal{G}$ and $\ket{g_k}$ is its corresponding eigenvector. We make an assumption that the eigenvalue $g_0 = 0$ to fix the ``origin'' of possible outcomes. 
Without defining the origin, we can make the expectation $\braket{\mathcal{G}}$ arbitrarily large while keeping its variance $\Delta \mathcal{G}$ constant, which in turn makes the fluctuation $\Delta\mathcal{G}^{2}/\braket{\mathcal{G}}^{2}$ arbitrarily small. If $\mathcal{G}$ is like a number operator, this condition is automatically satisfied. After the joint system goes through some unitary evolution, the projective measurement of $\mathcal{G}$ is performed on E. When we focus on the reduced description of S, we can express this entire process with the Kraus operators $\{ V_k \}$, which correspond to the projection onto the $k$th eigenspace of $\mathcal{G}$. Because each operator $V_k$ is defined with the $k$th eigenvector of $\mathcal{G}$ and the unitary operator, it is useful to formulate the fluctuation of $\mathcal{G}$ and the thermodynamic change that the whole system undergoes during the process. Finally, the minimal fluctuation of $\mathcal{G}$ meets the following TUR \cite{Hasegawa2021}:
\begin{equation}
  \dfrac{\Delta \mathcal{G}^2}{\langle \mathcal{G}\rangle^2} \geq \dfrac{1}{\Xi},
  \label{eq:opensys}
\end{equation}
where $\Xi$ is referred to as the survival activity, a generalization of a dynamical activity \cite{Maes2007, Baiesi2009, Maes2020}. The survival activity, which has a thermodynamic meaning, is concretely defined when $V_0^{\dagger}V_{0}$ is regular as follows:
\begin{equation}
  \Xi = \mathrm{tr}[ ( V_0^{\dagger} V_0)^{-1} \rho] -1 
  ,
  \label{eq:survivalactivity}
\end{equation}
where $\rho$ is the initial density operator of S. The definition Eq.~\eqref{eq:survivalactivity} is based on the two linked factors, $g_0 = 0$ and $V_0$, and lets us interpret the survival activity as a thermodynamic quantity that reflects the degree to which the system is excited against the outcome given by them. In particular, if we suppose classical Markov processes, we find $\Xi$ to be a dynamical activity with a sufficiently short duration \cite{Hasegawa2021}. The dynamical activity gives the expected number of jumping events, namely, quantifying the intensity of stochastic transitions of the system \cite{Maes2020}. We expect that, in such classical Markov processes, one can find the classical TUR bounded by the dynamical activity \cite{Hasegawa2021}.

Next, we consider a photon counting experiment and take $\mathcal{G}$ as a number operator again. 
The condition $g_0 = 0$ is the same as no photon detection, so we find $\mathrm{tr}[V_0^{\dagger}V_0 \rho]$, the probability of no event, by introducing the positive operator-valued measure $V_0^{\dagger}V_0$. Therefore, the expectation of the inverse of $V_0^{\dagger}V_0$ can be interpreted as the rarity of the event occurring within the monitored range. To make this intuition more clear, if the minimal eigenvalue of $V_0^{\dagger}V_0$ is small, $\Xi$ can have a large value, while $V_0^{\dagger}V_0 = I_\mathrm{S}$, meaning no transition from the origin, lets the survival activity converge to $0$. This implies that the survival activity can be regarded as a quantum generalization of a dynamical activity.

\section{\label{sec:4}Uncertainty Relation for Indirect Measurement}%

\subsection{Derivation}
In this section, we explain the derivation of our uncertainty relation for indirect measurement. 

Consider the principal system S and the attached probe system P, whose initial states are given by the density operators $\rho_{\mathrm{S}}$ and $\rho_{\mathrm{P}}$, respectively. We suppose that $\rho_{\mathrm{P}}$ has the spectral decomposition $\rho_{\mathrm{P}} = \sum_k q_k \ket{\phi_k}\bra{\phi_k}$, where $q_k$ is the probability of finding the eigenvector $\ket{\phi_k}$ with $\sum_k q_k = 1$. Here, we consider performing an indirect measurement on the combined system $\mathrm{S}+\mathrm{P}$, for the purpose of obtaining the mean value of the observable $A$ defined for S, which is given by $\langle A \rangle = \mathrm{tr}[A \rho_{\mathrm{S}}]$. Suppose that the meter observable $M$ for $\mathrm{S}+\mathrm{P}$ has the spectral decomposition $M = \sum_k r_k \Pi_k$, where the projection operator $\Pi_k$ gives rise to the outcome $r_k$ with $r_j \leq r_k$ for $j < k$. To fulfill the requirement of the proof of Eq.~\eqref{eq:opensys}, we set the eigenvalue $r_0$  as $0$ by shifting all eigenvalues. Of course, these conditions do not lose generality. We implement the unitary transform $U$, which is designed to satisfy the unbiasedness  condition that the projective measurement of $M$ can yield $\langle A\rangle$, as shown in Eq.~\eqref{eq:unbiased}. After measuring $M$ with its projectors $\{\Pi_k\}$ and tracing out the degrees of freedom of P, the post-measurement state $\rho_{\mathrm{S}}^\prime$ is obtained by
\begin{eqnarray}
  \rho^\prime_{\mathrm{S}} &=& \sum_k \mathrm{tr}_{\mathrm{P}}[ \Pi_k U ( \rho_{\mathrm{S}} \otimes \rho_{\mathrm{P}} ) U^{\dagger} \Pi_k ] \notag\\
  &=& \sum_{k,j,m} q_m \bra{\psi_j} \Pi_k U \ket{\phi_m} \rho_{\mathrm{S}} \bra{\phi_m} U^{\dagger} \Pi_k \ket{\psi_j} \notag\\
  &=:& \sum_{k,l} V_{k,l} \rho_{\mathrm{S}} V_{k,l}^{\dagger} \quad (l = (j,m)),
  \label{eq:post-measurement}
\end{eqnarray}
where $\{ \ket{\psi_j}\}$ is an arbitrary orthonormal basis of P and $\{ V_{k,l} \}$ are the Kraus operators satisfying the condition for the identity $I_\mathrm{S}$ with $\sum_{k,l} V_{k,l}^{\dagger}V_{k,l} = I_\mathrm{S}$.\ We can easily confirm that the mean value of the meter observable $M$ is also described by
\begin{equation}
  \langle M \rangle_{\mathrm{p}} = \sum_{k,l} r_k \mathrm{tr}[ V_{k,l}^{\dagger}V_{k,l} \rho_{\mathrm{S}}],
\end{equation}
where $\langle \cdots \rangle_{\mathrm{p}}$ is an ensemble of the post-measurement state and $\langle \cdots \rangle$ denotes the same but for the initial state. Through the formulation above, we can characterize the evolution of the indirect measurement when the initial density operator $\rho_{\mathrm{S}}$ and the Kraus operators $\{ V_{k,l} \}$ are provided.

\begin{figure}[!tb]
  \centering
  \includegraphics[scale=0.85, page=1]{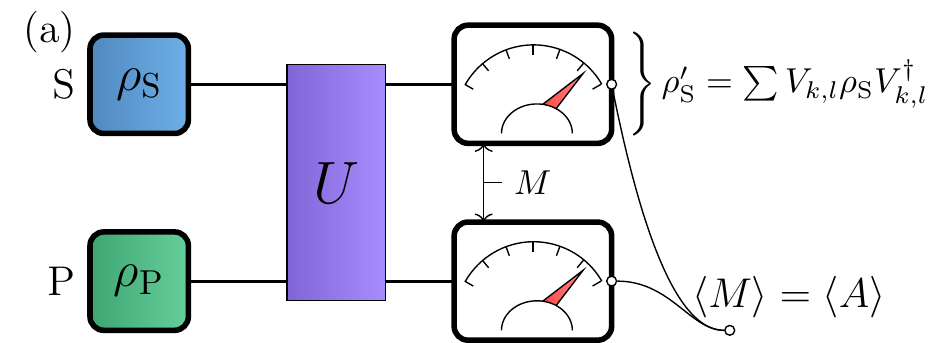}
  \includegraphics[scale=0.85, page=2]{fig1.pdf}
  \caption{Schematic illustration of the measurement setup. (a) Indirect measurement. To obtain the desired expectation $\langle A\rangle$ with the probe system P, one performs a projective measurement of $M$ on the entire system following the proper interaction given by $U$. (b) Installation equivalent to the indirect measurement (a) for the sake of applying the TUR. $\ket{\psi_{\mathrm{S}}}$ is the purified  state of $\rho_{\mathrm{S}}$ with the extended system $\mathrm{S}+\overline{\mathrm{S}}$. The ancillary system $\mathrm{Q}$ is introduced by following  Naimark's dilation. One can obtain the same outcome $\langle M\rangle$ by measuring $M^\prime$, which is constructed in Eq.~\eqref{eq:Mprime} in a way that $M^\prime$ contacts nothing but $\mathrm{Q}$.  The post-measurement state $\rho_{\mathrm{S}}^\prime$ is identical to one of the indirect measurements.}
  \label{fig}
\end{figure}

Next, we convert this configuration into a useful form, as illustrated in Fig.~\ref{fig}. We exploit the form of the post-measurement state $\rho_{\mathrm{S}}^\prime$ to evaluate the quantum Fisher information in the following proof of the uncertainty relation. This is achieved by associating the state $\rho_\mathrm{S}^\prime$ with the Stinespring representation. The state $\rho_{\mathrm{S}}^\prime$ can be also interpreted as being made up of the set of quantum operations $\{ \mathcal{E}_k\}$, where $\mathcal{E}_k(\cdot) = \sum_l V_{k,l}^{\dagger} (\cdot) V_{k,l}^{\dagger}$ and $\sum_k \mathcal{E}_k$ is trace-preserving. Each quantum operation $\mathcal{E}_k$ is specified by the subset $\{ V_{k,l} \}_l$; that is, the projector $\Pi_k$ appears in Eq.~\eqref{eq:post-measurement}. Namely, the ancillary systems $\overline{\mathrm{S}}, \mathrm{Q}$ and the isometric map $\mathcal{U}$ exist such that the pure state $\ket{\Psi}$ equivalent to $\rho_{\mathrm{S}}^\prime$ is given by \cite{Chuang2011}
\begin{eqnarray}
  \ket{\Psi} &=& \mathcal{U} (\ket{\psi_{\mathrm{S}}} \otimes \ket{e_0}) \notag \\
  &=& \mathcal{U} \left( \sum_{\mu} \sqrt{\lambda_{\mu}} \ket{\mu_{\mathrm{S}}} \otimes \ket{\mu_{\overline{\mathrm{S}}}} \otimes \ket{e_0}\right) \notag \\
  &=& \sum_{k,l} \sum_{\mu} \sqrt{\lambda_{\mu}} (V_{k,l} \ket{\mu_{\mathrm{S}}}) \otimes \ket{\mu_{\overline{\mathrm{S}}}} \otimes \ket{k,l}, \label{eq:purified}
\end{eqnarray}
where $\ket{\psi_{\mathrm{S}}}$ is the purified initial state of $\rho_{\mathrm{S}}$, whose spectral decomposition is $\rho_{\mathrm{S}} = \sum_{\mu} \lambda_{\mu} \ket{\mu_{\mathrm{S}}}\bra{\mu_{\mathrm{S}}}$; $\ket{\mu_{\overline{\mathrm{S}}}}$, which is defined on $\overline{\mathrm{S}}$, is a one-to-one state associated with $\ket{\mu_{\mathrm{S}}}$; $\ket{e_0}$ is an arbitrary pure state of $\mathrm{Q}$; and $\ket{k,l}$ are the unit vectors constituting the orthogonal basis of $\mathrm{Q}$. The vector $\ket{k,l}$ corresponds to the operator $V_{k,l}$ having an action on the system S in Eq.~\eqref{eq:purified}. By partially tracing out the density operator $\ket{\Psi}\bra{\Psi}$ for $\overline{\mathrm{S}}$ and $\mathrm{Q}$, we can recover the post-measurement state $\mathrm{tr}_{\mathrm{\overline{S},Q}^\prime}[ \ket{\Psi}\bra{\Psi}] = \sum_{k,l} V_{k,l} \rho_{\mathrm{S}} V_{k,l}^{\dagger}$, which agrees with the Kraus-Stinespring representation given by $\mathcal{U}$ and $\mathrm{Q}$.

The obtained state Eq.~\eqref{eq:purified} also provides another merit in that it lets us find  projection-valued measures (PVM) of the ancillary system $\mathrm{Q}$ apart from the principal system S. The indirect measurement scheme is allowed to perform the measurement on the total system, whereas the following proof has to consider the projective measurement limited to $\mathrm{Q}$. In the current situation, the projection operator $\Pi'_k$ is given by \cite{Chuang2011}
\begin{equation}
  \Pi_{k}' = \sum_{l} \ket{k,l} \bra{k,l},
\end{equation}
where the vectors $\{ \ket{k,l} \}$ are introduced in Eq.~\eqref{eq:purified}. This implies that a transformation applied to the composite system $\mathrm{S}+\overline{\mathrm{S}}+\mathrm{Q}$ resorts to Naimark's dilation theorem, which indicates that the PVM over the ancilla with $\Pi_{k}^\prime$ can take the place of a positive operator-valued measure with $\{ V_{k,l}\}$ \cite{Peres1990, Beneduci2020}. Finally, the indirect measurement started with the meter observable $M$ can be replaced with the counterpart observable defined on $\mathrm{Q}$:
\begin{equation}
  M' = \sum_k r_k \Pi_k' = \sum_{k,l} r_k \ket{k,l} \bra{k,l}.
  \label{eq:Mprime}
\end{equation}
We emphasize that this virtual operator works only in the ancillary system $\mathrm{Q}^\prime$. The counterpart $M^\prime$ can behave in the place of $M$ because the mean value satisfies
\begin{eqnarray}
  \langle M^\prime \rangle_{\mathrm{p}} &=& \braket{\Psi | I_{\mathrm{S,\overline{S}}} \otimes M^\prime | \Psi} \notag\\
  &=& \sum_{k,l} \sum_{k^\prime,l^\prime} r_k \braket{\psi_{\mathrm{S}} | V_{k,l}^{\dagger}V_{k^\prime,l^\prime} | \psi_{\mathrm{S}} } \braket{k,l | k^\prime,l^\prime} \notag\\
  &=& \sum_{k,l} r_k \braket{\psi_{\mathrm{S}} | V_{k,l}^{\dagger}V_{k,l} | \psi_{\mathrm{S}} } \notag\\
  &=& \langle M\rangle_{\mathrm{p}}, \label{eq:meanMR}
\end{eqnarray}
where $I_{\mathrm{S,\overline{S}}}$ is the identity of $\mathrm{S}+\overline{S}$ and also the variance is equivalent to the variance of $M$:
\begin{eqnarray}
  \Delta_{\mathrm{p}}M^{\prime 2} &=& \braket{\Psi | (I_{\mathrm{S,\overline{S}}} \otimes M^\prime)^2 | \Psi} - \braket{\Psi | I_{\mathrm{S,\overline{S}}} \otimes M^\prime | \Psi}^2 \notag \\
  &=& \sum_k r_k^2 \braket{\Psi | I_{\mathrm{S,\overline{S}}} \otimes \Pi_k^\prime | \Psi} - \langle M\rangle_{\mathrm{p}}^2 \notag\\
  &=& \sum_{k,l} r_k^2 \braket{\psi_{\mathrm{S}} | V_{k,l}^{\dagger} V_{k,l} | \psi_{\mathrm{S}}} - \langle M\rangle_{\mathrm{p}}^2\notag \\
  &=& \langle M^2 \rangle_{\mathrm{p}} - \langle M\rangle_{\mathrm{p}}^2 \notag \\
  &=& \Delta_{\mathrm{p}}M^2,
  \label{eq:varMR}
\end{eqnarray}
where $\Delta_{\mathrm{p}}(\cdots)$ denotes the standard deviation with respect to the post-measurement state. Therefore, we can transition from the original setting to this new setting given by the final state $\ket{\Psi}$ and the virtual operator $M^\prime$. 

We now go back to the thermodynamic uncertainty relation under these settings. Performing the projective measurement $M'$ and using the relation Eq.~\eqref{eq:opensys}, the bound is proved by the quantum Cram\'{e}r-Rao inequality \cite{Holevo2011, Hasegawa2018, Liu2020}. When the Kraus operators are parametrized as $V_{k,l}(\theta)$ with $\theta$ being a real value, the quantum Fisher information $\mathcal{I}_F(\theta)$ is bounded from above by $4(\braket{\Psi | H_1^2 |\Psi} - \braket{\Psi| H_2 |\Psi}^2)$ with $H_1 = \sum_{k,l} \partial_{\theta}V_{k,l}(\theta)^{\dagger} \partial_{\theta}V_{k,l}(\theta)$ and $H_2 = i\sum_{k,l} \partial_{\theta}V_{k,l}^{\dagger}(\theta)V_{k,l}(\theta)$ \cite{Escher2011}. Following the derivation of the TUR \cite{Hasegawa2021}, we parametrize the introduced Kraus operators $\{ V_{k,l}\}$ as follows:
\[
  V_{k,l}(\theta) = e^{\theta/2}V_{k,l},
\]
except for one arbitrary operator $V_{0,l}$. As mentioned in Sec.~\ref{sec:3}, this treatment is ascribed to the condition $r_0=0$, which fixes the origin of possible measurement outcomes. For the excluded one $V_{0,l}$, the condition $\sum_{k,l}V_{k,l}^{\dagger}(\theta) V_{k,l}(\theta) = I_{\mathrm{S}}$ leads to
\[
    V_{0,l}(\theta) = W\sqrt{I_{\mathrm{S}} - e^{\theta} \left(I_{\mathrm{S}} - V_{0,l}^{\dagger}V_{0,l} \right) },
\]
where $W$ is an arbitrary unitary operator. Here note that setting $\theta = 0$ recovers the original dynamics. Applying the Cram\'{e}r-Rao bound and taking $\theta \rightarrow 0$, we obtain
\[
  \forall l,\ \dfrac{\Delta_{\mathrm{p}}M^2 }{\langle M\rangle_{\mathrm{p}}^2} = \dfrac{\Delta_{\mathrm{p}}M^{\prime 2}}{\langle M^\prime\rangle_{\mathrm{p}}^2} \geq \dfrac{1}{\Xi_l}.
\]
We can admit the first equality by Eqs.~\eqref{eq:meanMR} and \eqref{eq:varMR}. We select the candidate to define the survival activity among the several Kraus operators $\{ V_{0,l} \}_l$, yielding the result $k=0$. Thus, the minimal survival activity can be chosen as
\begin{equation}
  \Xi = \min_{l} \{ \mathrm{tr}[ (V_{0,l}^{\dagger} V_{0,l})^{-1} \rho ] - 1\}
\end{equation}
for a meaningful bound. As a result, the bound is reduced to
\begin{equation}
  \dfrac{\Delta_{\mathrm{p}}M^2}{\langle M\rangle_{\mathrm{p}}^2} \geq \dfrac{1}{\Xi}
  \label{eq:tentative}
\end{equation}
for the indirect measurement of a single observable.

Based on the setting of the indirect measurement [see Eqs.~\eqref{eq:agreement_of_mean} and \eqref{eq:vardecomp}], the left-hand side of Eq.~\eqref{eq:tentative} can be written as
\[
  \dfrac{\Delta_{\mathrm{p}}M^2}{\langle M\rangle_{\mathrm{p}}^2} = \dfrac{\Delta A^2 + \Delta N^2}{\langle A\rangle^2}.
\]
Finally, we obtain
\begin{equation}
  \Xi \left(1 + \dfrac{\Delta N^2}{\Delta A^2}\right) \geq \dfrac{\langle A \rangle^2}{\Delta A^2}. \label{eq:finalresult}
\end{equation}
This result provides a trade-off relation for the indirect measurement. The left-hand side of Eq.~\eqref{eq:finalresult} represents the reciprocity between the survival activity $\Xi$ and the term concerned with the error $\Delta N$. The error term $1 + \Delta N^2 /\Delta A^2$ characterizes an additional amplitude caused by the extra error $\Delta N^2$ in accordance with the initial variance $\Delta A^2$. Then, the zero-noise $\Delta N$ means that an additional error is not increased by the indirect measurement. However, the right-hand side is a squared reciprocal of the CV $\Delta A/\langle A\rangle$. It is determined only by the initial state and the actual observable $A$, free from the measurement process; that is, the quantity acts as the ideal bound among feasible measurement processes.

This relation places importance on the thermodynamic aspect and the measurement process rather than the complementarity that the conventional measurement uncertainty relations have dealt with. In the joint measurement case, the noise operators, which correspond to their respective incompatible observables, come into question. The variances of these noise operators dependently exhibit how much the meter observables compensate the original observations. Unlike that dependency, the indirect measurement process, which our relation aims at, is not uniquely specified and is restricted in the form of Eq.~\eqref{eq:finalresult} in that more active dynamical processes can suppress an amplification of the error involved.

Here, we focus on the thermodynamic point of view. The entropy production of thermodynamic systems embodies the irreversibility of the undergoing process and dominantly restricts the precision of observables far from equilibrium. Compared with the entropy production, however, the dynamical activity that is defined as the average number of stochastic jumps works likewise but performs significantly when the system is instead near equilibrium. In other words, the kinetic aspect needs to be considered in thermodynamics. As discussed in Sec.~\ref{sec:3}, because the survival activity given by the underlying evolution can be seen as a generalization of the dynamical activity, we can find that the transition due to the indirect measurement deserves the thermodynamic contribution. This notion is reflected in the survival activity.

\subsection{Examples}

\begin{figure}[!tb]
  \centering
  \includegraphics[scale=0.6, page=1]{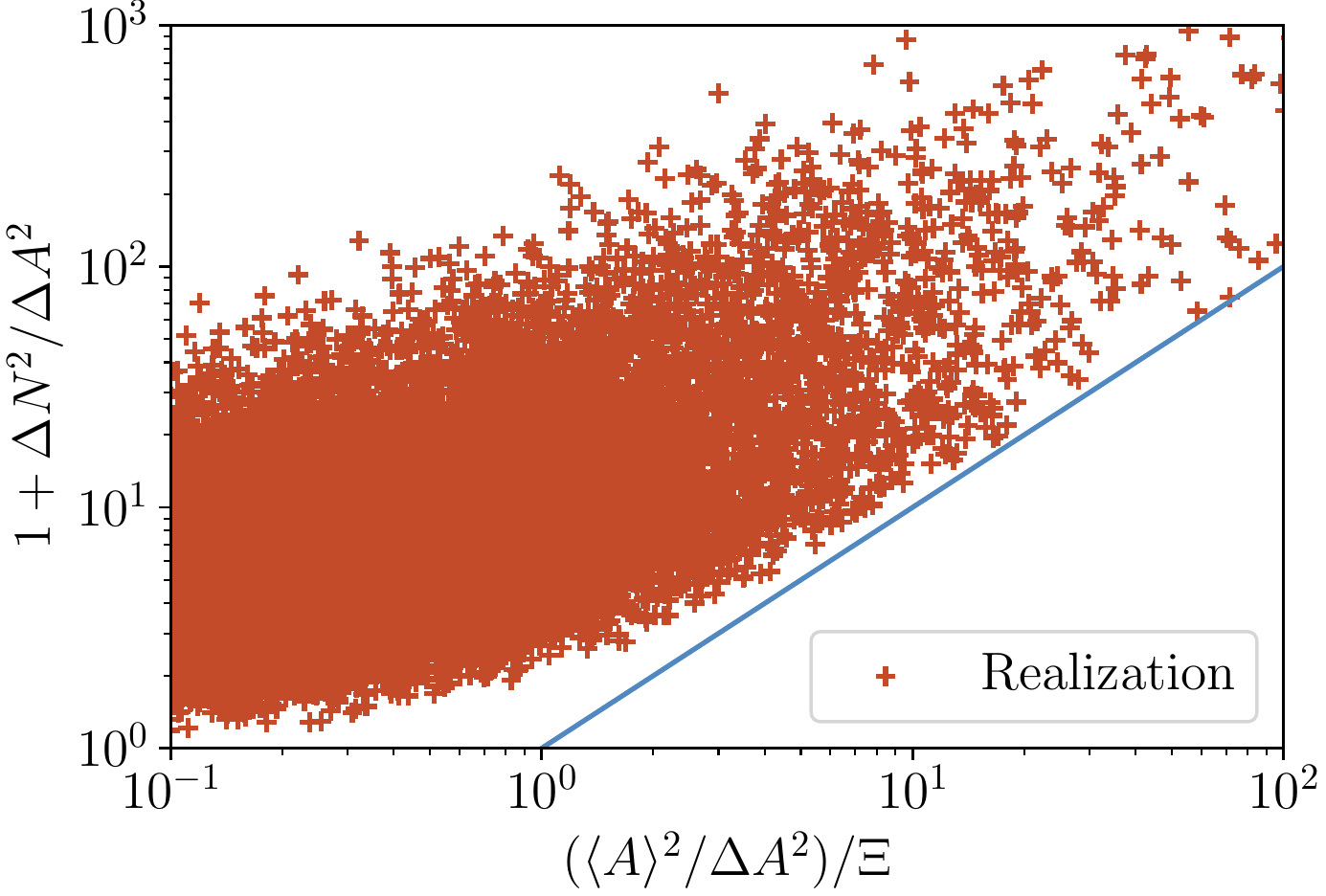}
  \caption{Numerical verification for the obtained bound Eq.~\eqref{eq:finalresult}. The dimensions of individual systems ($d_\mathrm{S}$ and $d_\mathrm{P}$) are randomly chosen from $\{2,3,4,5\}$ in each trial. For the probe's density operator $\rho_\mathrm{P}$, a unitary operator $U$ and a meter observable $M$, which are randomly generated in agreement with the requirement, are used to formulate the system observable $A$ under the unbiasedness condition. The brown crosses represent each calculation and the blue line is the obtained bound.}
  \label{fig:exp1}
\end{figure}

\begin{figure}[!tb]
  \centering
  \includegraphics[scale=0.6, page=1]{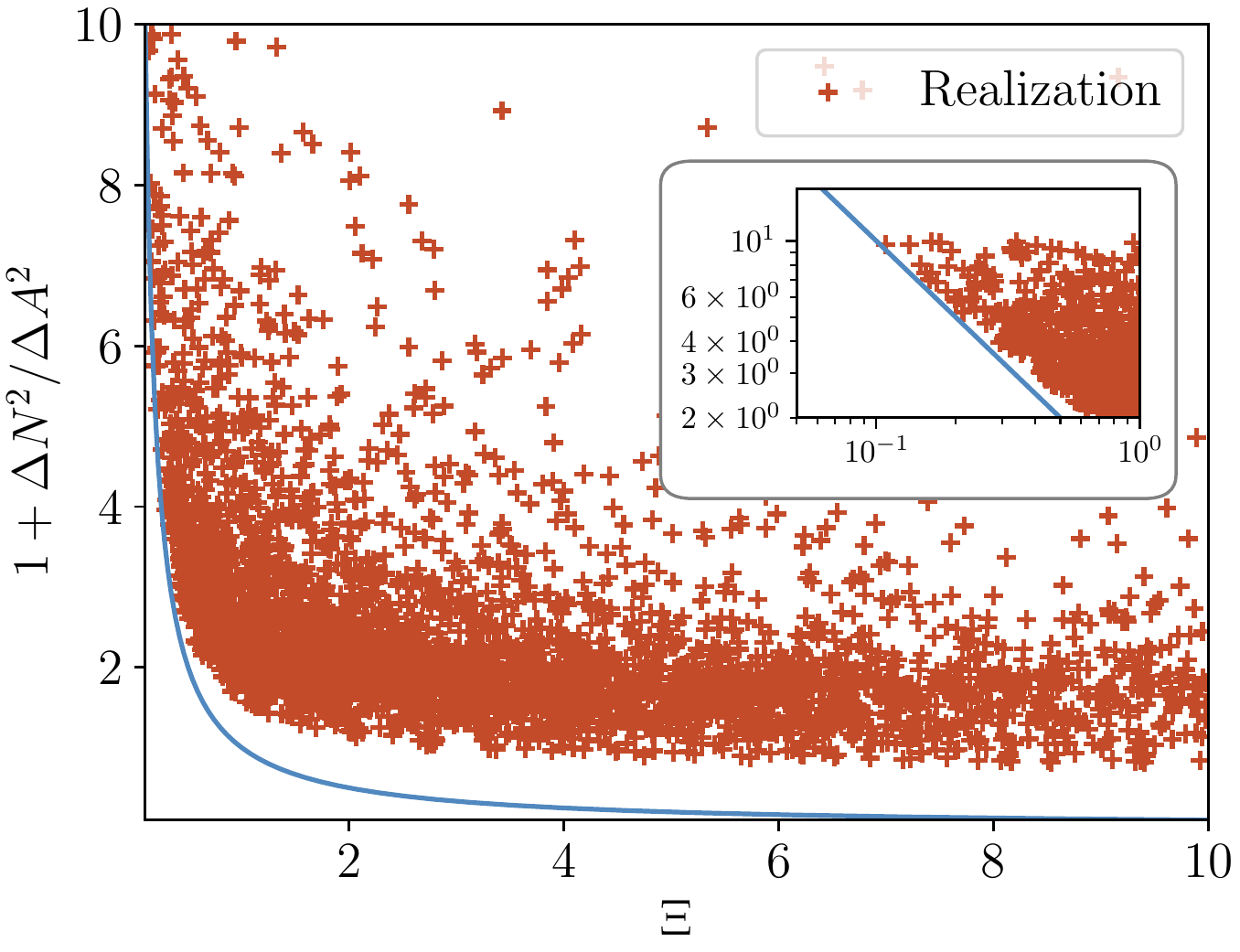}
  \caption{Trade-off relation of indirect measurement for a qubit system. To permit the coefficient of variation  $\langle A\rangle/\Delta A$ to be held constant, $A$ and $\rho_{\mathrm{P}}$ are fixed to Eq.~\eqref{eq:constants_exp2}. The $x$-axis takes the survival activity and the $y$-axis denotes $1 + \Delta N^2/\Delta A^2$. The inset shows the validity of our result again. Each indirect measurement procedure is numerically constructed with sufficiently small error $\| A - \mathrm{tr}_\mathrm{P}[U^{\dagger}(I_\mathrm{S}\otimes M)U(I_\mathrm{S}\otimes \rho_\mathrm{P})] \| \leq 10^{-5}$. The obtained data are scaled by dividing them by $\sqrt{\mathrm{CV}^2} = |\langle A\rangle|/\Delta A$.}
  \label{fig:exp2}
\end{figure}

To show the validity of the derived relation represented by Eq.~\eqref{eq:finalresult}, we performed two numerical experiments.

In the first numerical experiment, we let $d_\mathrm{S}$ and $d_\mathrm{P}$ be the dimensions of S and P, respectively, which are integers randomly selected from $2 \leq d_\mathrm{S}, d_\mathrm{P} \leq 5$. We also randomly generate a probe density operator $\rho_{\mathrm{P}}$, a unitary operator $U$, and a meter observable $M$. Due to the requirement for our setting, $M$ should have one or more zero eigenvalues. The observable $A$, which is the original observable of interest, is inversely determined through the unbiasedness condition Eq.~\eqref{eq:unbiased}; that is, we regard $\mathrm{tr}_{\mathrm{P}}[U^{\dagger}MU (I_{\mathrm{S}} \otimes \rho_\mathrm{P})]$ as equivalent to $A$.  Figure~\ref{fig:exp1} shows the result of the calculation, where the brown crosses denote each realization and the blue line represents the lower bound as a function of the squared CV $\langle A\rangle^2/\Delta A^2$ and the survival activity $\Xi$. This first case shows the validity of our obtained result with many randomized instances.

In the second numerical experiment, we focused on a more practical example that examines feasible procedures for indirect measurement. Let the dimensions be $d_\mathrm{S} = d_\mathrm{P} = 2$. In this qubit system, we fix the density operator of the principal system $\rho_{\mathrm{S}}$ and the observable $A$ as
\begin{equation}
  \rho_{\mathrm{S}} = \dfrac{1}{2}\left(\begin{array}{cc}1 & -i \\ i & 1\end{array}\right),\ A = \dfrac{1}{2}\sigma_z + I_{\mathrm{S}},
  \label{eq:constants_exp2}
\end{equation}
where $\sigma_z$ is the Pauli Z-operator. By setting $\rho_\mathrm{S}$ and $A$ beforehand as described above, the CV becomes constant. The meter observable $M$ is designed to act only on probe systems with eigenvalues of zero, namely $g_0=0$. Thus the identity operator $I_{\mathrm{S}}$ appears in $A$ so as to sustain the positivity inherited from such $M$ values. After we randomly select a unitary operator $U$ again, we numerically determine the probe's density operators $\rho_\mathrm{P}$ and elements of $M$ so that the unbiasedness condition can approximately hold with sufficient accuracy. Finally, these setups illustrate the trade-off between the indirect measurement procedures in Fig.~\ref{fig:exp2}. Note that the plotted values display $1 + \Delta N^2 / \Delta A^2$ as a function of $\Xi$ that is different from the previous case. The inset emphasizes the validity of the bound near saturation again. As seen in the figure, this case implies that a smaller activity unavoidably gives rise to larger noise.

\subsection{Noise-Disturbance Scenario}

\begin{figure}[!tb]
  \centering
  \includegraphics[scale=1.5, page=1]{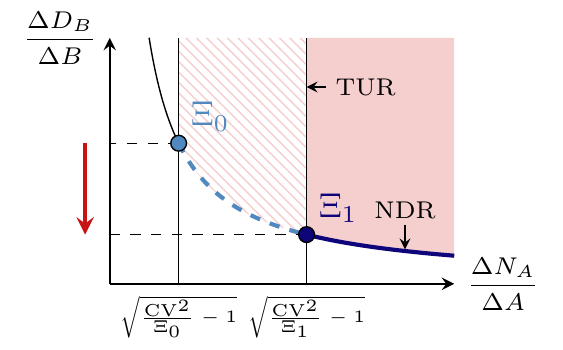}
  \caption{Illustration of the effect of the thermodynamic uncertainty on the noise-disturbance uncertainty. The curve indicated by the NDR is a lower bound of the error-disturbance uncertainty relations given by Eq.~\eqref{eq:noise-disturbance2}. The vertical lines and shaded areas annotated with TUR denote the obtained relation that gives a constraint on $\Delta N_A$ with the survival activity $\Xi$. When one moves $\Xi_0$ to $\Xi_1$ (here $\Xi_0 > \Xi_1$, shown as blue dots), the optimal strategy, including $\Delta D_B$, is affected. Our relation implies that increasing $\Delta D_B$ cannot decrease $\Delta N_A$ at some point.}
  \label{fig:edur}
\end{figure}

The joint measurement, which underlies the measurement uncertainty relations, can achieve precise measurement for incompatible observables in an approximate way. Even if one supposes the existence of meter observables that commute with each other, we will simply obtain the same relation as Eq.~\eqref{eq:finalresult}. Meanwhile, in the following noise-disturbance scenario, we can rather find a different perspective.

Consider an indirect measurement where the meter observable is defined on the probe system. We use the meter observable $M_A$ of the probe to approximate the measurement of $A$. Thus, the noise operator is given by $N_A = U^{\dagger}(I_{\mathrm{S}} \otimes M_A) U - A \otimes I_{\mathrm{P}}$, where $U$ is a unitary operator that implements a certain interaction. Under the above setting, we can also evaluate the effect on the incompatible observables due to indirect measurement. Let the observable $B$ of the system satisfy $[A, B] \neq 0$. Then, the disturbance operator of $B$ brought along with the measurement of $M_A$ is defined as
\begin{equation}
  D_B := U^{\dagger} (B \otimes I_{\mathrm{P}}) U - B \otimes I_{\mathrm{P}}.
  \label{eq:disturbance}
\end{equation}
That is, the disturbance operator $D_B$ quantifies recoil on $B$ due to the indirect measurement by simply considering the difference between before and after the interaction. The concept of disturbance is essential for retaining the original distribution during quantum non-demolition measurements. Its magnitude is described by the variance of the disturbance operator $\Delta D_B^2$, which is analogous to the noise operator.

Heisenberg's argument about the measurement of position and its effect on momentum has been a seminal idea for the noise-disturbance relation (NDR) \cite{Heisenberg1927}, but it can be violated in some circumstances \cite{Erhart2012}. In contrast, according to Ozawa \cite{Ozawa2003}, we have the strictly valid NDR:
\begin{equation}
  \Delta N_A \Delta D_B + \Delta N_A \Delta B + \Delta A \Delta D_B \geq \dfrac{1}{2}| \langle [A, B]\rangle |,
  \label{eq:noise-disturbance}
\end{equation}
where the noise $\Delta N_B$ is replaced with the disturbance $\Delta D_B$. Our result and the NDR of Eq.~\eqref{eq:noise-disturbance} can have a thermodynamic effect on the disturbance $\Delta D_B$. For clarity, the NDR of Eq.~\eqref{eq:noise-disturbance} is rewritten as the following reciprocal relation:
\begin{equation}
  \left(\dfrac{\Delta N_A}{\Delta A} + 1\right)\left(\dfrac{\Delta D_B}{\Delta B} + 1\right) \geq 1 + \dfrac{| \langle [A,B] \rangle|}{ 2\Delta A \Delta B}.
  \label{eq:noise-disturbance2}
\end{equation}
This expression leads us to interpret Ozawa's NDR as the trade-off between the amplitudes of the noise $\Delta N_A$ and the disturbance $\Delta D_B$ in the form of a standard deviation. At this point, due to Eq.~\eqref{eq:noise-disturbance2}, our result also implies that a survival activity of the whole system intermediately places a restriction on the disturbance of the incompatible counterpart by limiting the feasible region of $\Delta N_A$. This idea is depicted in Fig.~\ref{fig:edur}. Suppose that one tries to implement an indirect measurement while considering a survival activity. Simply based on the NDR, one can reduce the noise $\Delta N_A$ by sacrificing disturbance $\Delta D_B$ along the lower bound. In regard to a thermodynamic perspective, our result, which here we say is TUR compared with NDR, argues that, for a sufficiently small survival activity $\Xi$, the noise $\Delta N_A$ has a non-trivial lower bound $\sqrt{\mathrm{CV}^2/\Xi - 1}$. The optimal strategy, which is given by the NDR, partially loses its freedom of feasibility because of a possible $\Delta N_A$ that depends on our TUR. As shown in Fig.~\ref{fig:edur}, if we decrease $\Xi_0$ to $\Xi_1$, a larger $\Delta D_B$ no longer contributes to suppressing $\Delta N_A$. Therefore, to deal with noise, the system needs to tolerate not only a certain disturbance but also a certain activity. That is, a dynamically active measurement process can implicitly support the utility of disturbance. 

\section{Conclusion}

To further explore recently revealed  thermodynamic uncertainty relations, the thermodynamic insight into quantum measurements was investigated. Measurement uncertainty has been proven with the indirect measurement method, which is one of the practical quantum measurement methods, and hitherto elucidated the complementarity of quantum measurement. The finding that particular thermodynamic quantities play a role in controlling the minimal precision of observables was used to derive distinct constraints on quantum measurement by taking advantage of the meter observable. That is, it was revealed that indirect measurement has a trade-off relation between its additional noise and the survival activity that quantifies the degree of exposure of the quantum system to stochastic transitions. Our result can be connected to the noise-disturbance relation in a thermodynamic sense again and lead to limiting the disturbance of an incomparable observable.

\begin{acknowledgments}
  This work was supported by KAKENHI Grant Numbers JP19K12153 and JP22H03659.
\end{acknowledgments}

\bibliography{bibliography}

\end{document}